\documentstyle[12pt]{article}

\begin{document}

\begin{center}

{}~\vfill

{\large \bf  Gauge and parametrization dependencies  \protect \\
of the one-loop counterterms in the Einstein gravity}

\vfill

{\large M. Yu. Kalmykov}
\footnote { E-mail: $kalmykov@thsun1.jinr.dubna.su$
 \protect \\
Supported in part by ISF grant \# RFL000}

\vspace{2cm}

{\em Bogoliubov Laboratory of Theoretical Physics,
 Joint Institute for Nuclear  Research,
 $141~980$ Dubna $($Moscow Region$)$, Russian Federation}

\end{center}

\vfill

\begin{abstract}

The parametrization and gauge dependencies of the one-loop counterterms
on the mass-shell in the Einstein gravity are investigated.
The physical meaning of the loop calculation results on the mass shell
and the parametrization dependence of the renormgroup functions
in the nonrenormalizable theories are discussed.
\end{abstract}

\vfill

PACS number  0460, 1110, 1115

\pagebreak

\section{Introduction}

The construction of quantum gravity is one of the fundamental
problems in the modern theoretical physics. The coincidence of the
Einstein gravity with all experimental results is very
good~\cite{will}; however, this theory is not renormalizable.
Einstein's gravity is the finite theory at the one-loop level in the
absence of both matter fields and cosmological
constant~\cite{THV}. But this theory diverges in the two-loop
order~\cite{GS},~\cite{V}.  The interaction of the gravity with
the matter fields gives rise to nonrenormalizable theories yet at the
one loop level ~\cite{DPN1},~\cite{DPN3}. Attempts to improve
the renormalizability of this theory by adding the terms
quadratic in the curvature tensor or corresponding matter fields
failed. In the first case, the obtained theory is
renormalizable but it is not unitary because the ghosts and
tachyons are present in the spectrum of the theory~\cite{St2} -
{}~\cite{John}.  The second case led to the discovery of supergravity
{}~\cite{SUGRA1},~\cite{SUGRA2}.  Due to the presence of the local
supersymmetry, supergravity is two-loop finite. But, the divergent
terms are present in the three-loop order~\cite{SUGRA3},
{}~\cite{SUGRA4}. Recently, the superstring ~\cite{GSW}
and the canonical approach ~\cite{A1} to quantum gravity have been
proposed as a sensible theory of quantum gravity.

Most of the calculations confirming the perturbation
nonrenormalizability of quantum gravity have been made by
the background field method ~\cite{BDW-67} - ~\cite{Arev}. This
method was suggested to obtain  covariant
results of the loop calculations. In the background field method, all
dynamical fields $\varphi^j $ are expanded with respect to background
values, according to $$ \varphi^j = \varphi^j_b  + \phi^j_q, $$ and only
the quantum fields $\phi^j_q$ are integrated over in the path integral.
The background fields $\varphi^j_b$ are effectively external
sources.  For the one-particle irreducible diagrams there is a
difference between the normal field theory and the background field
method insofar as the gauge-fixing term may introduce additional
vertices.  B.DeWitt has proved that these additional vertices do not
influence the $S$-matrix and the $S$-matrix in the formalism of the
background field method is equivalent to the conventional $S$-matrix
{}~\cite{BDW-67},~\cite{BDW}.  This proof has later been extended
in a lot of papers ~\cite{TH1} - ~\cite{vor3}. Hence, the counterterms
on the mass-shell in the background field method must be independent of
the gauge-fixing parameters and the reparametrization of the quantum
fields. These statements are called the DeWitt-Kallosh theorem
{}~\cite{BDW-67},~\cite{BDW},~\cite{Kal} and equivalence theorem
{}~\cite{equiv1} - ~\cite{equiv4} respectively.

The equivalence theorem states, that the $S$-matrix of the
renormalizable theory is independent of the following change of
variables:

$$
\varphi^j \rightarrow '\varphi^j = \varphi^j +
\left( \varphi^2 \right)^j + \left( \varphi^3 \right)^j + \dots
$$

In the case of the quantum gravity, this statement is divided into two
parts:

\begin{enumerate}
\item It is well know that there is considerable freedom in what one
considers to be gravitational fields. We can consider the arbitrary
tensor density  $\tilde{g}_{\mu \nu } = g_{\mu \nu }(-g)^r$ or
$\tilde{g}^{\mu \nu } = g^{\mu \nu }(-g)^s$ as gravitational
variables. In accordance with the equivalence theorem the loop
counterterms on the mass-shell must be independent of the choices of
gravitational variables.
\item The loop counterterms on the mass-shell are independent of the
redefinition of quantum fields of the form
$$
h_{\mu \nu } \rightarrow 'h_{\mu \nu }  = h_{\mu \nu }  +
{\it k} \left( h^2 \right)_{\mu \nu } +
{\it k}^2 \left( h^3 \right)_{\mu \nu } + \dots
$$
This redefinition must influence
only the higher loop results off the mass-shell.
\end{enumerate}

By means of the corresponding choice of  gravitational variables or
the corresponding quantum field redefinition, one can considerably
reduce the number and the type of interaction vertices. For example, if
we consider $g_{\mu \nu }$ as a gravitational variable, the number of
three-point interactions in the Einstein gravity is equal to
13~\cite{GS}; if the tensor density $g^{\mu \nu } \sqrt{-g}$ is
selected as a dynamical variable, the number of three-point
interactions is equal to six~\cite{CLR}; combining both method reduces
the number of three-point interactions to two~\cite{V}. However, the
Einstein gravity is not a renormalizable theory.  Thus, the equivalence
theorem may not be fulfilled. A systematic search in the present
context has never been undertaken in quantum gravity. The dependence of
the nonlinear renormalizable quantum field theories on the choice of
the parametrization have been investigated in the paper ~\cite{tyutin}.

The DeWitt-Kallosh theorem asserts that the loop counterterms on the
mass-shell calculated by means of the background field method are
independent of the gauge-fixing term. This statement has been verified
in many papers ~\cite{GPN} - ~\cite{Keon}.  It turns out that the
proof of this theorem is valid only for renormalizable theories
(such as Yang-Mills theory, QED, QCD). For nonrenormalizable theories
(such as gravity) the proof of the DeWitt-Kallosh theorem is formal.
For example, it has recently been was shown that the one-loop
counterterms of the Einstein gravity on the mass-shell depend on the
gauge fixing terms~\cite{Ichi1}.  Moreover, we can choose a gauge so
that the Einstein gravity interacting with the matter fields will be
renormalizable at the one-loop level.

Because of the complexity of the gauge suggested in papers
{}~\cite{Ichi1}, one needs to create a new algorithm for obtaining
covariant one-loop results.  It will be very nice to verify the gauge
dependencies of the one-loop counterterms in a simpler gauge by
using the standard well defined algorithm ~\cite{Gil} - ~\cite{Bar}.

The main goal of the present paper is to investigate the influence of
the field parametrization and the gauge fixing term on the one-loop
counterterms of the Einstein gravity on the mass-shell.

We use the following notation:
$$ c = \hbar = 1;~~~~~ \mu , \nu  = 0,1,2,3;~~~~~ {\it k}^2 = 16 \pi
G,~~~~~(g) = det(g_{\mu \nu }), ~~~~~\varepsilon  = \frac{4-d}{2} $$
$$ R^\sigma _{~\lambda  \mu  \nu } = \partial_\mu \Gamma^\sigma
_{~\lambda \nu }  - \partial_\nu \Gamma^\sigma _{\lambda \mu } +
\Gamma^\sigma_{~\alpha \mu } \Gamma^\alpha_{~\lambda  \nu } -
\Gamma^\sigma_{~\alpha  \nu }  \Gamma^\alpha_{\lambda \mu }~~~~~,
R_{\mu \nu } = R^\sigma_{~\mu \sigma \nu },~~~~~
  R =  R_{~\mu \nu } g^{\mu \nu } $$
where $\Gamma^\sigma_{~\mu \nu } $ is the Riemann connection.

Objects marked by the tilde $\tilde{} $  are the tensor densities.
The other are the tensors.

\section{One-loop counterterms}

One considers the Einstein gravity with the cosmological constant. The
action of the theory is

\begin{equation}
{\it S}_{gr}  =  - \frac{1}{{\it k}^2} \int d^4 x \sqrt{-g}
\left(R  - 2 \Lambda \right)
\label{action}
\end{equation}

For calculating the one-loop counterterms we use the background field
method and the Schwinger-DeWitt technique. In the gauge theories, the
renormalization procedure may violate the gauge invariance at the
quantum level, thus destroying the renormalizability of the theory.
Therefore, one is bound to apply an invariant renormalization. This can
be achieved by applying an invariant regularization and using the
minimal subtraction scheme ~\cite{min},~\cite{Sp}. It has been proved
that the dimensional regularization ~\cite{dim1} - ~\cite{dim4} is an
invariant regularization preserving all the symmetries of the classical
action which do not depend explicitly on the space-time dimension
\cite{Sp},~\cite{action},~\cite{vlad}.  It has been shown ~\cite{AKK}
that in general renormalizable and nonrenormalizable theories the
background field formalism requires using an invariant
renormalization procedure to  obtain valid results.  A
noninvariant regularization or renormalization may break an implicit
correlation between different diagrams, which is essential as one
formally expands the action in the background and quantum fields. We
will use the invariant regularization (dimensional renormalization and
minimal subtraction scheme) in our calculations.

When using the invariant renormalization the one-loop correction to
the usual effective action is

\begin{equation}
\Gamma^{(1)}  = \frac{i}{2} \left( \ln {\it det} \triangle_{ab} -
2 \ln {\it det} \triangle_{FP} \right)
\label{one-loop}
\end{equation}

where

$\triangle_{FP}$ is the Faddeev-Popov ghost operator;

\begin{equation}
\triangle_{ab}  = \frac{\delta^2 S(\phi)}{\delta \phi^a \delta
\phi^b} + P^j_{~a}(\phi ) P_{jb}(\phi )
\label{def}
\end{equation}

and $P^j_{~a}(\phi )$ is a gauge fixing term.

The divergence part of the one-loop effective action  obtained by
means of the heat kernel method is

\begin{equation}
\Gamma^{(1)}_\infty  = -\frac{1}{32 \pi^2 \varepsilon }
\int d^4 x \sqrt{-g} \bigl( B_4(\triangle_{ab}) - 2 B_4(\triangle_{FP})
\bigr)
\label{div}
\end{equation}
\noindent where  $B_4$  is the second coefficient of the
spectral expansion of the corresponding differential operator. For the
operator

\begin{equation}
\triangle_{ij} = - \left(  \nabla^2 {\bf 1}_{ij} + 2 S^\sigma_{~ij}
\nabla_\sigma + X_{ij} \right)
\label{oper}
\end{equation}

$B_4$ is equal to

\begin{equation}
B_4(\triangle) = {\it Tr} \Biggl( \frac{1}{180}
\left( R^2_{\mu \nu \sigma \lambda } - R^2_{\mu \nu } \right)
+ \frac{1}{2} \left( Z + \frac{R}{6} \right) + \frac{1}{12}
Y_{\mu \nu } Y^{\mu \nu } \Biggr)
\label{B}
\end{equation}

where

\begin{eqnarray}
Z & = & X - \nabla_\lambda S^\lambda - S_\lambda S^\lambda  \nonumber \\
Y_{\mu \nu }  & = & \nabla_\mu S_\nu - \nabla_\nu S_\mu  +
S_\mu S_\nu  - S_\nu S_\mu  + [\nabla_\mu , \nabla_\nu ]{\bf 1}
\end{eqnarray}

In the general case the dynamical variable in some
metrical theory of gravity is the tensor density $\tilde{g}_{\mu
\nu } = g_{\mu \nu }(-g)^r$ or $\tilde{g}^{\mu \nu } = g^{\mu \nu
}(-g)^s$, where $r$ and $s$ are the numbers satisfying the
conditions

\begin{equation}
{\it det} \left|
\frac{\partial \tilde{g}_{\mu \nu }}{\partial g_{\alpha \beta }}
\right| \neq 0
\label{cond1}
\end{equation}

or

\begin{equation}
{\it det} \left|
\frac{\partial \tilde{g}^{\mu \nu }}{\partial g^{\alpha \beta }}
\right| \neq 0
\label{cond2}
\end{equation}

In accordance with the background field method we rewrite the dynamical
field as

\begin{equation}
\tilde{g}_{\mu \nu } = \tilde{g}_{\mu \nu } + {\it k}
\tilde{h}_{\mu \nu }
\label{exp}
\end{equation}

where $\tilde{g}_{\mu \nu }$ is the classical part satisfying
the following equation

\begin{equation}
\frac{\delta {\it S}_{gr}}{\delta \tilde g^{\mu \nu}}  =
 R_{\alpha \beta}  - \frac{1}{2} R g_{\alpha \beta}
\frac{2r + 1}{4r + 1} + \Lambda g_{\alpha \beta}
\frac{1}{4r + 1}   = 0
\label{mass-shell1}
\end{equation}

Having solved this equation we obtain

\begin{equation}
R_{\mu \nu} = \Lambda g_{\mu \nu}
\label{mass-shell2}
\end{equation}

Some functions of $g_{\mu \nu }$ and their expansion in powers of
the quantum field are given below

\begin{eqnarray}
(-g) & = & (-\tilde g)^{\frac{1}{t}} \nonumber \\
g_{\mu \nu } & = & \tilde g_{\mu \nu }  (- \tilde g)^{-\frac{r}{t}}
\nonumber \\
g^{\mu \nu } & = & \tilde g^{\mu \nu } (- \tilde g)^{\frac{r}{t}}
\nonumber \\
\Gamma^\sigma_{~\mu \nu} & = & \frac{1}{2} \tilde g^{\sigma \lambda}
\biggl( - \partial_\lambda \tilde g_{\mu \nu}  +
\partial_\mu \tilde g_{\lambda \nu} +
\partial_\nu \tilde g_{\mu \lambda} \biggr)
-  \frac{r}{2t}~  \tilde g^{\alpha \beta}
\partial_\lambda \tilde g_{\alpha \beta}
\biggl( \delta^\sigma_\mu \delta^\lambda_\nu
+ \delta^\sigma_\nu \delta^\lambda_\mu
- \tilde g^{\sigma \lambda} \tilde g_{\mu \nu} \biggr)
\nonumber \\
(-g)^m & = & (-g)^m \left( 1 + {\it k} \frac{m}{t}h +
 \frac{{\it k}^2}{2}\left( \frac{m^2}{t^2} h^2  - \frac{m}{t} h_{\alpha
\beta } h^{\alpha \beta } \right)  + O({\it k}^3) \right)
\nonumber \\
g^{\mu \nu } & = & g^{\mu \nu } + {\it k}
\left( \frac{r}{t} h g^{\mu \nu }  - h^{\mu \nu } \right) +
{\it k}^2 h^{\mu \alpha }h^\nu _{~\alpha }
\nonumber \\
& + & {\it k}^2 \left( \frac{r^2}{2 t^2} h^2  - \frac{r}{2t} h_{\alpha \beta }
h^{\alpha \beta } \right) g^{\mu \nu }  - {\it k}^2 \frac{r}{t}
h h^{\mu \nu }  + O({\it k}^3)
\nonumber \\
g_{\mu \nu } & = & g_{\mu \nu } - {\it k}
\left( \frac{r}{t} h g^{\mu \nu }  - h^{\mu \nu } \right) +
 {\it k}^2 \left( \frac{r^2}{2 t^2} h^2  + \frac{r}{2t} h_{\alpha
\beta } h^{\alpha \beta } \right) g_{\mu \nu }  - {\it k}^2 \frac{r}{t}
h h_{\mu \nu }  + O({\it k}^3)
\nonumber \\
\Gamma^\sigma_{~ \mu\nu} & = & \Gamma^\sigma_{~ \mu \nu} +
 \frac{{\it k}}{2} g^{\sigma \lambda}
 \biggl(\nabla_\mu h_{\nu \lambda} + \nabla_\nu h_{\mu \lambda} -
 \nabla_\lambda h_{\mu \nu} \biggr)
\nonumber \\
&  -  &  \frac{{\it k}}{2}  \left( \frac{r}{t} \right) g^{\alpha \beta}
 \nabla_\lambda h_{\alpha \beta} \biggl(
 \delta^\lambda_\mu \delta^\sigma_\nu + \delta^\lambda_\nu
\delta^\sigma_\mu - g^{\sigma \lambda} g_{\mu \nu} \biggr)
\nonumber \\
& - & \frac{{\it k}^2}{2} h^{\lambda \sigma } \left(
-\nabla_\lambda h_{\mu \nu }  + \nabla_\mu h_{\nu \lambda } +
\nabla_\nu h_{\mu \lambda }  \right)
\nonumber \\
& + & \frac{{\it k}^2}{2} \frac{r}{t} h^{\alpha \beta } \nabla_\lambda
h_{\alpha \beta } \left(\delta^\sigma_\mu  \delta^\lambda_\nu
+ \delta^\sigma_\nu  \delta^\lambda_\mu   -
g^{\sigma \lambda } g_{\mu \nu } \right) + O({\it k}^3)
\end{eqnarray}

where
\begin{equation}
h = h_{\alpha \beta } g^{\alpha \beta }
\label{h}
\end{equation}

\begin{equation}
 t \equiv 4r + 1 \neq 0
\label{t}
\end{equation}

We expand the action (\ref{action}) in powers of the quantum field and
pick out the terms quadratic in the quantum fields

\begin{eqnarray}
{\it L}_{eff} & = &  \Biggl(
\frac{1}{4} \nabla_\sigma h_{\alpha \beta}
\nabla_\lambda h^{\alpha \beta} g^{\sigma \lambda} -
\frac{1}{2} \nabla_\mu h^\mu_{~\nu} \nabla_\lambda h^{\lambda \nu}
+ \frac{1}{2}~ \frac{2r + 1}{t} \nabla_\mu h \nabla_\nu h^{\mu \nu}
\nonumber \\
& - & \frac{1}{2}~ \frac{6r^2 + 4r + 1}{2 t^2}  \nabla_\sigma h
\nabla_\lambda h g^{\sigma \lambda}  - \frac{1}{2} h^{\mu \nu}
X_{\mu \nu \alpha \beta} h^{\alpha \beta} \Biggr) \sqrt{-g}
\label{effective}
\end{eqnarray}
where
\begin{eqnarray}
X_{\mu \nu \alpha \beta} & = & R_{\mu \alpha} g_{\nu \beta}
+ R_{\mu \alpha \nu \beta} - 2 \rho R_{\mu \nu} g_{\alpha \beta} +
\rho^2 R g_{\mu \nu} g_{\alpha \beta}
\nonumber \\
& - & \rho R g_{\mu \alpha } g_{\nu \beta}
+ \frac{\Lambda}{t}~ g_{\mu \alpha} g_{\nu \beta}
- \frac{\Lambda}{2 t^2}~ g_{\mu \nu} g_{\alpha \beta}
\label{X}
\end{eqnarray}

\begin{equation}
\rho  \equiv \frac{t + 1}{4t}
\end{equation}

The effective Lagrangian (\ref{effective}) is invariant under the
general coordinate transformation

\begin{eqnarray}
x^\mu  \rightarrow  'x^\mu & = & x^\mu +
{\it k} \xi^\mu(x) \nonumber \\
\tilde h_{\mu \nu}(x)    \rightarrow   '\tilde h_{\mu \nu}(x) & = &
\tilde h_{\mu \nu}(x) - {\it k} \partial_\mu \xi^\alpha \tilde
g_{\alpha \nu}(x) - {\it k} \partial_\nu \xi^\alpha \tilde g_{\mu
\alpha}(x) \nonumber \\
& - & {\it k} \xi^\alpha \partial_\alpha \tilde g_{\mu
\nu}(x)   - 2r {\it k} \partial_\alpha \xi^\alpha \tilde g_{\mu \nu}(x)
+ O({\it k}^2)
\label{coordin}
\end{eqnarray}

Now we investigate the parametrization dependence of the one-loop
counterterms. To use the standard method of calculation
(\ref{oper}) and (\ref{B}), we fix the gauge invariance by the
following condition:

\begin{equation}
F_\mu  = \nabla_\nu \tilde h^\nu_{~\mu } - \rho \nabla_\mu \tilde h
\label{gauge1}
\end{equation}

\begin{equation}
{\it L}_{gf} = \frac{1}{2} F_\mu F_\nu g^{\mu \nu }
(-g)^{\frac{1-2r}{2}}
\label{gauge2}
\end{equation}

The ghost action obtained in the standard way is

\begin{equation}
{\it L}_{gh}  = \overline{c}^\mu \left(
g_{\mu \nu } \nabla^2  + R_{\mu \nu } \right) c^\nu \sqrt{-g}
\label{ghost}
\end{equation}

The one-loop counterterms off the mass-shell are

\begin{eqnarray}
\triangle \Gamma^{(1)}_{\infty} & =&  - \frac{1}{32 \pi^2 \varepsilon}
\int d^4x \sqrt{-g}  \Biggl( \Lambda^2 \biggl( 8 + 2 t^2 - 8 t +
\frac{18}{t^2} \biggr)
\nonumber \\
& + & \Lambda R \biggl(
- \frac{4}{3} - t^2 + \frac{8}{3}t - \frac{9}{t^2} \biggr) +
R_{\mu \nu} R^{\mu \nu}  \biggl( - \frac{3}{10} + 2 t - t^2 \biggr)
\nonumber \\
& + & \frac{53}{45} \biggl( R_{\mu \nu \sigma \lambda}
R^{\mu \nu \sigma \lambda} - 4 R_{\mu \nu } R^{\mu \nu } + R^2 \biggr)
\nonumber \\
& + &
   R^2 \biggl(- \frac{49}{60}  + \frac{3t^2}{8}  - \frac{2t}{3} +
\frac{9}{8t^2} \biggr) \Biggr)
\label{result}
\end{eqnarray}

On the mass-shell we have

\begin{equation}
\triangle \Gamma^{(1)}_{\infty} = - \frac{1}{32 \pi^2 \varepsilon}
\int d^4x \sqrt{-g}  \Biggl(\frac{53}{45} R_{\mu \nu \sigma \lambda}^2
- \frac{58}{5} \Lambda^2 \Biggr)
\label{on-result}
\end{equation}

This result coincides with the result obtained in the paper
{}~\cite{ChD1}.  The one-loop counterterms on the mass-shell calculated
in the gauge (\ref{gauge1}) in the Einstein gravity are independent of
the choice of  parametrization of the gravitational field.  The case
$\tilde g^{\mu \nu } = g^{\mu \nu }(-g)^s$ can be considered
analogously way and does not give essentialy new results.

Now, we change the gauge fixing term  and investigate the
gauge  and parametrization dependencies of the one-loop counterterms on
the mass-shell.  The most general gauge linear in the quantum field is

\begin{equation}
F_\mu  = \alpha \nabla_\nu \tilde h^\nu_{~\mu } + \beta  \nabla_\mu
\tilde h  + T_{\mu \sigma \lambda } \tilde h^{\sigma \lambda }
+  S_\mu^{~\nu \sigma \lambda } \nabla_\nu \tilde h_{\sigma
\lambda }
\label{gengauge}
\end{equation}
where

$\alpha$ and $\beta $ are the arbitrary constants;

$T_{\mu \alpha \beta }$ and $S_\mu^{~\nu \alpha \beta }$ are some
tensors depending on the background field $g_{\mu \nu }$, functions of
$g_{\mu \nu }$ (such as $R^\sigma_{~\alpha \lambda \beta }, R_{\mu \nu
}, R$) and the covariant derivatives $\nabla_\sigma $. Expression
(\ref{gengauge}) being the most general gauge for the gravity, is
defined by the following conditions:

\begin{enumerate}
\item Lorentz covariance
\item the number of  derivatives with respect to the quantum fields is
smaller than or equal to one
\item linear in the quantum field
\end{enumerate}

Using a gauge of this type, one can simplify the calculations
of counterterms in some models.

In the previous papers, the one-loop counterterms for
the Einstein gravity were calculated in the following gauges:

\begin{enumerate}
\item  $r = 0; T_{\mu \alpha \beta } = S_\mu^{~\nu \alpha \beta } = 0$
{}~\cite{KTT},~\cite{CD3}

The one-loop counterterms off the mass-shell depend on the
parameters  $\alpha$ and $\beta $. On the mass the one-loop counterterms
coincide with the result (\ref{on-result})

\item  $r = 0; \alpha = \beta = 0; S_\mu^{~\nu \alpha \beta } = 0$
{}~\cite{L}

The calculations were made by means of the diagrams technique. The
metric was expanded around the flat background. It is impossible to write
the results of calculations  in the covariant way.

\item $r = 0; T_{\sigma \mu \nu } = 0$ ~\cite{Ichi1}

To calculate the one-loop counterterms in the covariant way,
one needs to create a new algorithm for the calculations. The
results on the mass-shell depend on the $S_\mu^{~\nu  \alpha
\beta }$
\end{enumerate}

We consider the case $r \neq 0, ~\alpha = 1, ~\beta = -\rho ,
S_{\mu }^{~\nu \alpha \beta } = 0, T_{\sigma \mu \nu } \neq 0$

This choice of parameters allows us to use the standard algorithm
for the one-loop calculations. We will use this gauge for
investigation of gauge and parametrization dependencies of the one-loop
counterterms on the mass-shell. The gauge involved is the following

\begin{equation}
F_\mu  = \nabla_\nu \tilde h^\nu_{~\mu } - \rho \nabla_\mu \tilde h +
T_{\mu \nu \sigma } \tilde h^{\nu \sigma }
\label{our-gauge}
\end{equation}

The arbitrary tensor $U_{\sigma \mu \nu }$ can be
decomposed into its irreducible parts:

\begin{equation}
U_{\sigma \mu \nu }  = A_\sigma g_{\mu \nu }  + B_\mu g_{\nu \sigma }
+ C_\nu g_{\mu \sigma }  + \frac{1}{6}
\varepsilon_{\sigma \mu \nu \lambda }  \check{U}^\lambda  +
\overline{U}_{\sigma \mu \nu }
\end{equation}
\noindent where $\check{U}^\lambda$ is the axial part defined by

\begin{equation}
\check{U}^\lambda  = \varepsilon^{\lambda \sigma \mu \nu }
U_{\sigma \mu \nu }
\end{equation}
\noindent and $A_\sigma ,B_\mu and C_\nu ,$ are the vector fields
defined by

\begin{equation}
A_\sigma  \equiv \frac{1}{18} \left( 5U_{\sigma \lambda }^{~~\lambda }
- U^\lambda_{~\sigma \lambda } - U^\lambda_{~\lambda \sigma }  \right)
\end{equation}

\begin{equation}
B_\sigma  \equiv \frac{1}{18} \left( -  U_{\sigma \lambda }^{~~\lambda
 } + 5 U^\lambda_{~\sigma \lambda } - U^\lambda_{~\lambda \sigma }
\right)
\end{equation}

\begin{equation}
C_\sigma  \equiv \frac{1}{18} \left( - U_{\sigma \lambda }^{~~\lambda }
- U^\lambda_{~\sigma \lambda } + 5 U^\lambda_{~\lambda \sigma }
\right)
\end{equation}
\noindent
and $\overline{U}_{\sigma \mu \nu }$ is the traceless part satisfying
the following conditions:

\begin{equation}
\overline{U}^\nu_{~\mu \nu }  = \overline{U}^\nu_{~\nu \mu }
= \overline{U}_{\nu \mu }^{~~\mu } \equiv 0
\end{equation}

\begin{equation}
\overline{U}_{\sigma \mu \nu } + \overline{U}_{\nu \sigma \mu } +
\overline{U}_{\mu \nu \sigma } = 0
\end{equation}

The tensor $T_{\sigma \mu \nu }$ presented in the gauge
(\ref{our-gauge}) satisfies the condition

\begin{equation}
T_{\sigma \mu \nu } = T_{\sigma \nu \mu }
\end{equation}

Then, the decomposition of $T_{\sigma \mu \nu }$ can be written in
the following way:

\begin{equation}
T_{\sigma \mu \nu }  = T_\sigma g_{\mu \nu } +
C_\nu g_{\mu \sigma } + C_\mu g_{\nu \sigma }  +
\overline{T}_{\sigma (\mu \nu )}
\end{equation}

The number of counterterms off the mass shell including $T^4$ and
$RT^2$, where $T^4$ and $RT^2$  are the symbolic notation for
contractions of the tensor $T_{\sigma \mu \nu }$ or the curvature tensor
and tensor $T_{\sigma \mu \nu }$, respectively, are about $150$. The
calculation of these counterterms is very cumbersome. To
reduce the number of possible counterterms and to facilitate the
calculations, we consider the three particular cases of the gauge
(\ref{our-gauge}).

\begin{enumerate}
\item $T_{\sigma \mu \nu }$ is an arbitrary tensor satisfying two
conditions
\begin{itemize}
\item $T_{\sigma \mu \nu }$ is the symmetrical tensor:
$T_{\sigma \mu \nu } = T_{(\sigma \mu \nu )}$
\item $T_{\sigma \mu \nu }$ is the traceless tensor:
$T_{(\sigma \mu \nu )}g^{\mu \nu } = 0$
\end{itemize}
\item $T_{\sigma \mu \nu } = T_\sigma g_{\mu \nu }$

where $T_\sigma$ is an arbitrary vector.
\item $T_{\sigma \mu \nu } = C_\mu g_{\sigma \nu }  + C_\nu g_{\sigma
\mu }$
where $C_\sigma$ is an arbitrary vector.
\end{enumerate}

In the first case, the gauge fixing term is

\begin{equation}
F_\mu  = \nabla_\nu \tilde h^\nu_{~\mu }  - \rho \nabla_\mu \tilde h +
T_{(\mu \alpha \beta )} \tilde h^{\alpha \beta }
\label{g1}
\end{equation}

The ghost action is

\begin{equation}
{\it L}_{gh}  = \overline{c}^\mu \left( g_{\mu \nu } \nabla^2  +
2 T^\sigma_{~\mu \nu } \nabla_\sigma  + R_{\mu \nu } \right) c^\nu
\sqrt{-g}
\label{gh1}
\end{equation}

In the second case, the gauge fixing term and the ghost action are:

\begin{equation}
F_\mu  = \nabla_\nu \tilde h^\nu_{~\mu }  - \rho \nabla_\mu \tilde h +
T_\mu \tilde h
\label{g2}
\end{equation}

\begin{equation}
{\it L}_{gh}  = \overline{c}^\mu \left( g_{\mu \nu } \nabla^2  +
2 t T_\mu \nabla_\nu  + R_{\mu \nu } \right) c^\nu \sqrt{-g}
\label{gh2}
\end{equation}

In the third case, the gauge fixing term and the ghost action are:

\begin{equation}
F_\mu  = \nabla_\nu \tilde h^\nu_{~\mu }  - \rho \nabla_\mu \tilde h +
2C_\nu  \tilde h^\nu_{~\mu }
\label{g3}
\end{equation}

\begin{equation}
{\it L}_{gh}  = \overline{c}^\mu \left( g_{\mu \nu } \nabla^2  +
2 C_\nu  \nabla_\mu + 2g_{\mu \nu } C^\sigma \nabla_\sigma +
4r C_\mu \nabla_\nu + R_{\mu \nu } \right) c^\nu \sqrt{-g}
\label{gh3}
\end{equation}

The results of the one-loop calculation on the mass-shell coincide with
the standard results (\ref{on-result}). Hence the one-loop counterterms
of the Einstein gravity on the mass-shell do not depend on the
choice of the tensor  $T_{\sigma \mu \nu }$.

\section{The physical meaning of the results of the loop calculation}

It is well known that in quantum gravity the
results of the loop calculations off the mass shell calculated by
means of the background field method depend on the gauge fixing term
and the choice of the parametrization of quantum fields. For example,
one considers the result (\ref{result}) and using the standard arguments
calculates some renormgroup functions. One considers only first two
counterterms ($\Lambda^2 $ and $\Lambda R$). The expression $\int d^4x
\sqrt{-g}\left( R_{\mu \nu \sigma \lambda } R^{\mu \nu \sigma \lambda }
- 4 R_{\mu \nu } R^{\mu \nu } + R^2 \right)$ is proportional to the
topological number of space-time, the so called Euler number. Hence,
this expression is some number. It can be assumed that other structures
appearing in one-loop counterterms ( $R^2_{\mu \nu }$ and $R^2$) are
comparably small in concrete physical applications. This situation
corresponds to the low energy limit.
 From the renormalization group analysis \cite{rg}, \cite{smolin} it is
well known that the terms with higher derivatives play the essential
role only in the high energy limit. But in the low energy limit the
essential role belongs to the terms with two derivatives. In this way,
in the low energy limit we consider only the $\Lambda^2$ and $R\Lambda$
terms.  Then, under this consideration the theory is renormalizable. At
the one loop level, one needs to renormalize the cosmological constant
$\Lambda $ and the gravitational constant ${\it k}^2$.
The cosmological constant can be represented in the following form:

\begin{equation}
\Lambda  = \frac{\lambda }{{\it k}^2}
\end{equation}
\noindent where $\lambda $ is the dimensionless constant. Then, from
expression (\ref{result}) one gets the renormalization group equations

\begin{equation}
\beta_\lambda  = \mu^2
\frac{\partial \overline{\lambda }}{\partial \mu^2 }   =
- \frac{\overline{\lambda}^2}{32 \pi^2 }
\left( (t-2)^2 + \frac{9}{t^2}\right)
\label{res2a}
\end{equation}

\begin{equation}
\gamma_{{\it k}^2}  = \mu^2
\frac{\partial \overline{{\it k}^2}}{\partial \mu^2 }
=  - \frac{\overline{ \lambda} }{32 \pi^2 }\overline {{\it k}}^2
\left( t^2 - \frac{8}{3}t +\frac{4}{3} + \frac{9}{t^2} \right)
\label{res2b}
\end{equation}
\noindent where $\mu^2 $  is the renormalization point mass and $\gamma$
is the anomalous dimension of the gravitational constant ${\it k}^2$. We
see that asymptotical freedom for the cosmological constant $\lambda$
is preserved for an arbitrary choice of the field parametrization. But
anomalous dimension of the gravitational constant ${\it k}^2$
drastically depends on the parametrization. In general, it is possible
to find such a parametrization that the  anomalous dimension will be
equal to zero. The parametrization dependence of the renormalization
group functions, such as the $\beta $-function and anomalous dimension,
have the same treatment as the gauge and scheme dependencies of these
functions in the ordinary quantum field theory.

In general, in the nonrenormalizable quantum gravity  all numerical
coefficients of the counterterms  calculated by means of the background
field method off the mass-shell depend on the choice of the gauge
and parametrization. The standard choice of $g_{\mu \nu }$ and
$\Phi_{mat}$ as dynamical variables, where $g_{\mu \nu }$ and
$\Phi_{mat} $ are the metric and material fields, respectively, is
simply a particular choice of a possible parametrization. The loop
counterterms off the mass shell obtained by means of these variables
are also parametrization dependent.

The dependence of the one-loop counterterms off the mass-shell on the
method of calculation have been discussed also in paper
{}~\cite{shapiro}. It has been shown that the Einstein gravity in the
first-order formalism corresponds to some choice of the parametrization
of the metric field.

In this situation the question arises: what is the physical
parametrization?

Quite recently Fujikawa has suggested a very beautiful way to
define the physical parametrization ~\cite{F1}. The true
dynamical variables are defined from the anomaly-free condition on the
BRST-transformation connected with the general coordinate
transformations. This prescription must be fulfilled for each
variable separately. This condition means that the dynamical variables
are  some tensor densities $\tilde{\varphi }$ obtained from the initial
fields $\varphi $ by multiplication by corresponding degree of $(-g)$
For example, the physical dynamical variables in the quantum gravity
must be $g_{\mu \nu }(-g)^{\frac{N-4}{4N}}$ or
$g^{\mu \nu }(-g)^{\frac{N+4}{4N}}$ where $N$ is the
space-time dimension. All material fields must be  replaced by some
tensor density fields. These results can be obtained from the
functional integral approach ~\cite{F2} without the BRST-symmetry.
However, the gauge dependence of the results of the loop calculation
is present even in this physical parametrization.

An other way to obtain physical results is the use of the
gauge and parametrization independent Vilkovisky-DeWitt effective action
instead of the ordinary effective action ~\cite{VDW1} -~\cite{BOS}.
But the calculations of the loop correction to the Vilkovisky-DeWitt
effective action are very cumbersome because of the nonlocal terms
in its definition. Moreover, the gauge and parametrization
invariance has been proved only for the renormalizable theories. In
the nonrenormalizable theories the Vilkovisky-DeWitt effective action
can give rise to gauge or parametrization dependent results off the
mass shell. The connection between ordinary and Vilkovisky-DeWitt
effective actions in the Einstein gravity have been discussed in paper
{}~\cite{LOT}.

To summarize, in nonrenormalizable theories the results of the loop
calculations off the mass shell within the background
field method  are physically meaningless.  For the
results of the loop background field method calculations on the mass
shell in some nonrenormalizable theory are be physically meaningful, one
needs to prove or  verify the validity of the DeWitt-Kallosh theorem and
equivalence theorem for this theory.

\section{Conclusion}

The background field formalism is a powerful tool for the loop
calculations. Its validity is based on the statement that the S-matrix
in the formalism of the background field method is equivalent to the
conventional S-matrix. The consequence of this equivalence is the gauge
and parametrization independence of the loop counterterms on the mass
shell calculated in the background field method.
For nonrenormalizable theories, such as Einstein gravity
the proof of this statement is formal.
In this way the question arises about the physical meaning of the loop
results calculated by the background field method
in the nonrenormalizable theories. Can we obtain some physical quantities
or some physical information from these calculations? If the DeWitt-Kallosh
theorem and equivalence theorem are fulfilled in some nonrenormalizable
theory, then it is possible to obtain some physical information from the
results of the loop calculations on the mass shell. Therefore, one needs
to verify the validity of the DeWitt-Kallosh theorem and equivalence
theorem for each nonrenormalizable theory.

In this paper the gauge and parametrization
dependencies of the one-loop counterterms of the Einstein gravity were
verified. The gauge (\ref{our-gauge}) and arbitrary
parametrization were considered. It turns out that on the mass shell
the one-loop counterterms do not depend on the considered gauge and
parametrization.  However, as has been shown in papers ~\cite{Ichi1},
the one loop counterterms on the mass shell in the most general gauge
(\ref{gengauge}) depend on the gauge parameter.  Hence, the
DeWitt-Kallosh theorem is not valid in this gauge.

What is the reason?. Maybe one needs to modify the statement of the
DeWitt-Kallosh theorem for nonrenormalizable gauge theories?. For
example, we can say that in the nonrenormalizable gauge theories the
DeWitt-Kallosh theorem is valid only in the physical, so called
Landau-DeWitt gauge, defined as

\begin{equation}
f_a = {\it R}_{a \beta }(\varphi ) \phi^\beta
\end{equation}
\noindent
where $\varphi^\alpha$ and $\phi^\alpha  $ are the background  and
quantum fields, respectively, and ${\it R}_{i \alpha }(\varphi )$ are
the generators of the gauge transformations. For the quantum gravity,
the Landau-DeWitt gauge is defined by

\begin{equation}
f_\mu  = \nabla_\nu h^\nu_{~\mu }  + \beta \nabla_\mu h
\end{equation}

\begin{equation}
L_{gf}  = \frac{1}{2\alpha } f_\mu f^\mu
\end{equation}
\noindent where $\alpha $ and $\beta $  are arbitrary numbers. In
papers ~\cite{KTT},~\cite{CD3} it has been shown that the one-loop
counterterms on the mass shell do not depend on the gauge parameters
$\alpha $ and $\beta $.  Then, we suggested that in the
Landau-DeWitt gauge the effective action would be connected with the
S-matrix.  Hence, the results obtained in the Landau-DeWitt gauge on
the mass-shell have the physical meaning.  Then, in the gauge distinct
>from Landau-DeWitt gauge the ordinary effective action on the
mass-shell does not imply physical quantities and one needs to define
some reduction method to obtain physical quantities from the usual
effective action in a nonphysical gauge.

To verify this statement one needs to calculate the gauge
dependence of the one-loop counterterms on the mass shell in the gauge
distinct from the Landau-DeWitt gauge. The gauge (\ref{our-gauge})
and gauge (\ref{gengauge})  satisfy
this condition. These gauges are equivalent. Then, the results of the
loop calculations in these gauges must have the same physical meaning.
In the gauge (\ref{gengauge}) the one-loop counterterms on the mass
shell depend on the gauge fixing term and, as consequence, are
meaningless. Hence, the results of the loop calculations in the
gauge (\ref{our-gauge})  do not have physical meaning as well. But the
results of the loop calculations in the gauge (\ref{our-gauge}) on the
mass-shell coincide with the results of the loop calculations in the
standard gauge (\ref{gauge1}). Then, the results of calculations by the
loop background field method in the Einstein gravity in
an arbitrary gauge do not have the physical meaning. We cannot obtain
some physical information from these calculations. In this way the
results of the loop calculations do not give information about
renormalizability of the theory.

It is possible that in  arbitrary nonrenormalizable
theories the ordinary effective action (and maybe the
Vilkovisky-DeWitt effective action)  on the mass-shell does not give
any physical quantities at all. The validity of the DeWitt-Kallosh and
equivalence theorem in particular gauges does not contradict
this statement.  This is simply a fortunate event. To
obtained physical information in nonrenormalizable theories, one
needs to define the physical quantities and to calculate loop
corrections only to these physical quantities.

I am very grateful to L.V.Avdeev, D.I.Kazakov and B.L.Voronov for many
usefull discussions. I am greatly indebted to G.Sandukovskaya for
critical reading of the manuscript.

\end{document}